\documentstyle[epsfig]{aipproc}

\begin{document}
\title{\vskip-1.5cm{\large\hfill LU TP 99--33\\\hfill October 1999\\
\hfill hep-ph/9910415\\[0.5cm]}Models of Low Energy
Effective Theory\\applied to Kaon Non-leptonic
Decays\\and Other Matrix Elements\thanks{Work
supported in part by TMR, EC-Contract No. ERBFMRX-CT980169
(EURO\-DA$\Phi$NE).}$^,$\thanks{To appear in the proceedings of the
  International
  Workshop on Hadron Physics `Effective Theories of Low Energy QCD', Coimbra,
  Portugal, September 10-15, 1999}}

\author{Johan Bijnens$^*$}
\address{$^*$Department of Theoretical Physics 2,
Lund University\\
S\"olvegatan 14A, S 22362 Lund, Sweden}
\maketitle

\begin{abstract}
In this talk I describe work on computing non-leptonic matrix elements
consistently with both long and short distance contributions included.
On the simpler example of the $\pi^+$-$\pi^0$ mass difference I explain
in detail the matching procedure and the difference between various low-energy
models. I then explain the new difficulties in non-leptonic Kaon decays
and how the matching here can in principle be done in the same way when
scheme dependences are correctly accounted for. In the end I summarize
the results J.~Prades and I obtain for the $\Delta I=1/2$ rule and $B_6$.
\end{abstract}

\section*{Introduction}

The problem of describing non-leptonic decays is a very old one and is still not
fully solved today. In this talk I will describe the large $N_c$ method
first suggested in a series of papers by Bardeen, Buras and G\'erard\cite{BBG}
and later extended by several other authors.
I will illustrate most of the problems and solutions on the example
of the charged and neutral pion mass difference and afterwards
show how this method can be extended systematically to the
case of non-leptonic
weak decays as well. The main results described there are those of \cite{BP1}
and also descibed in \cite{talks}.

The subject of this meeting was hadronic physics, so why are we interested
in these extra quantities. They provide a very strong test of our
understanding of the strong interaction at all length scales.
Our present knowledge of the strong interaction can be summarized as:
\begin{description}
\item[Short Distance:] This is the perturbative QCD domain and here
QCD has had many successes, we count this region as understood.
\item[(Very) Long Distance:] This is the Chiral Perturbation
Theory (CHPT) regime\cite{CHPTreview}.
Many successes again and basically understood.
\item[Intermediate Distance:] This is the domain of models supplemented
with various arguments, sum rules, lattice QCD results, etc. and
is the most difficult.
\end{description}
In the type of observables covered in this talk all three regimes are
important. We consider processes with incoming and outgoing hadrons
but with an internally exchanged photon or weak boson.
The difficulty now resides in the fact that even if the external hadrons
have all low momenta we need to integrate over all momenta of the internal
$\gamma$ or $W^+$. This means that all regimes come into play and that they
need to be connected properly to each other. The last is known as matching.

The main part is in Sect. \ref{pippi0} where I show how we can explain the
mass difference, $m_{\pi^+}^2-m_{\pi^0}^2$ using this class of methods.
Here we can also see how the model approach and the correct answer agree.
Sect. \ref{kaon} then covers the extra problems involved in non-leptonic weak
decays and how the $X$-boson method of \cite{BP1} can be used to solve those.
Finally I present numerical results for this case and conclusions.

\section{A simple example: The $\pi^+$-$\pi^0$ mass difference.}
\label{pippi0}

This non-leptonic matrix element has several features that make it
simpler.
\begin{enumerate}
\item We can neglect $m_u$ and $m_d$ to a rather good approximation. This then
allows current algebra to relate the electromagnetic mass difference
to a vacuum to vacuum matrix element only\cite{das}. This can then be related
to the measured hadronic cross-sections in electron-positron annihilation so
in this case we know the correct answer.
\item There are no large masses involved so there are no large
logarithms that need resummation.
\item The photon itself provides for an easy identification of correct scales.
\end{enumerate}

Basically the procedure is now to evaluate
\begin{equation}
\label{massdiff}
m^2_{em} = -\langle M|e^2\int \frac{d^4q}{(2\pi)^4}
\frac{J_\mu(q)J_\nu(-q)}{q^2}\left(g_{\mu\nu}-\xi\frac{q_\mu q_\nu}{q^2}
\right)|M\rangle\,.
\end{equation}
where $M$ stands for the meson under consideration and $J_\mu$ for the
electromagnetic current. $\xi$ is a gauge parameter.
The procedure is now as follows:\\
{\bf 1:}We rotate the integral over photon momenta in Eq. (\ref{massdiff})
to Euclidean space. This has two advantages, in Euclidean space thresholds
and poles are smoothed out making treatment of these easier and Euclidean
space momenta have all components small if $q^2_E$ is small. The latter
allows for a simpler identification of long and short-distance than in
Minkowski space.\\
{\bf 2:}
The final step is now to set
\begin{equation}
\int{d^4q_E} = \int_0^\infty q_E^3dq_E\int d\Omega = 
\underbrace{
\int_0^\mu q_E^3dq_E\int d\Omega}_{\mbox{long-distance}}
+\underbrace{\int_\mu^\infty q_E^3dq_E\int d\Omega}_{\mbox{short-distance}}
\end{equation}
and perform both integrals separately. Notice that the scale $\mu$ is
just a splitting scale in the integral and is not directly related
to any subtraction scale in the calculation itself. Therefore, if both 
the long-distance (from 0 to $\mu$) and the short-distance are calculated
with high enough precision the final result should be independent of the value
of $\mu$. We check this by varying $\mu$ in all our calculations, i.e. we check
the matching.
 
\subsection{Short-Distance}

\begin{figure}[b!] 
\centerline{\epsfig{file=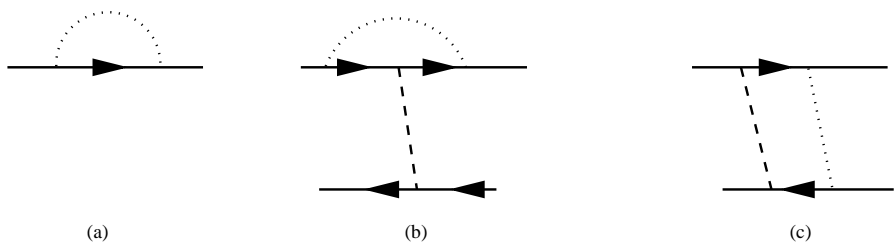,width=0.8\textwidth}}
\vspace{10pt}
\caption{The three short-distance contributions, (a) electromagnetic quark-mass
corrections. (b) Penguin-like diagrams (c) Box Diagrams.
The dashed line is a gluon, the dotted line a photon and the full lines
are quarks.}
\label{figdashen}
\end{figure}

The short-distance contribution was first calculated in \cite{BBGpi}
using the sum rule by Das et al.\cite{das}. It was later rederived
using the Operator Product expansion in \cite{dashen}. The diagrams in
Fig.~\ref{figdashen} depict the main contributions.
Performing the photon integral leads to a set of four-quark operators
that can be evaluated in leading $1/N_c$ since we can then apply
factorization.
The result is \cite{BBGpi,dashen}
\begin{equation}
\left(m_{\pi^+}^2-m_{\pi^0}^2\right)_{SD}
=\frac{3\alpha_S \alpha}{\mu^2}F^2 B_0^2 =
\frac{3\alpha_S \alpha}{\mu^2}\frac{\langle\bar{q}q\rangle^2}{F^2}\,,
\end{equation}
with $F$ the pion decay constant in the chiral limit and $B_0$
the parameter in lowest order CHPT describing the quark condensate.

\subsection{Long-Distance}

In the previous subsection we could use perturbative QCD but that is not
possible in the long distance domain. So here we have to put in the things we
know and try various models.
\begin{description}
\item[CHPT:] This can be done for $\mu$ rather small and leads to
\begin{equation}
\left(m_{\pi^+}^2-m_{\pi^0}^2\right)_{LD}
= \frac{3\alpha}{4\pi}\left(\mu^2 + \frac{2 L_{10}}{F^2}\mu^4\right)\,.
\end{equation}
The first term was first done in \cite{BBGpi} and the chiral correction
in \cite{Bijnens,Knecht1}. The two contributing diagrams are depicted
in Fig. \ref{figdashen2}.
\begin{figure}[t!] 
\centerline{\epsfig{file=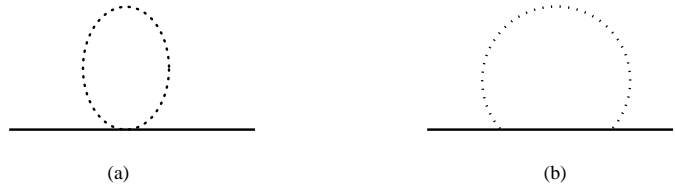,width=0.6\textwidth}}
\vspace{10pt}
\caption{The long-distance contributions to 
$\left(m_{\pi^+}^2-m_{\pi^0}^2\right)$.
The dotted line is a photon and the full lines
are pions.}
\label{figdashen2}
\end{figure}
Something that is important is that the gauge dependence only cancels between
the two diagrams in Fig.~\ref{figdashen2}.
\item[Chiral Quark Model:] This was done in \cite{BdR} and gives only a
marginal
improvement. Note that we cannot use the usual dimensional regularization
here but must use the cut-off in the photon propagator. There is the additional
problem that at first sight only the equivalent of the diagram of
Fig~\ref{figdashen2}(a) appears, which is a two-loop diagram,
 and the result is not gauge invariant.
Only after the equivalent of (b) is added, which is a three-loop diagram,
does the gauge dependence cancel as required\cite{BdR}.
\item[With Vector-axial-vector Mesons:]
We have to include here Weinberg's constraint on the couplings to
obtain a unique result otherwise the result will be very dependent
on the specific model used. E.g. a hidden gauge model with only vector
mesons is still quadratic in $\mu^2$ but with a negative coefficient.
Using Weinberg's constraints leads to
\begin{equation}
\label{VMD}
\left(m_{\pi^+}^2-m_{\pi^0}^2\right)_{LD} =
\frac{3\alpha}{2\pi}M_V^2\log\left[
\frac{M_V^2+\mu^2}{M_A^2+\mu^2}\frac{M_A^2}{M_V^2}\right].
\end{equation}
But beware of partial results. Using a linear vector
representation only even gave a quartic dependence on $\mu$\cite{Bijnens}.
The result in (\ref{VMD}) for $\mu\to\infty$ is basically the
result of \cite{das} and was also obtained in \cite{Ecker89}.
It has also several nice features. Expanding in $\mu$ for small $\mu$
reproduces the CHPT result with the meson dominated value for $L_{10}$.
For large  $\mu$ it goes as $1/\mu^2$ so it can match on very well
to the earlier short-distance result.
\item[ENJL] This basically coincides with the previous result and was
first obtained in \cite{BRZ}.
\end{description}
Extensions of the above exists for nonzero-quark masses\cite{dashen,BP2}
and references therein and also with more large $N_c$ arguments to
underpin the matching\cite{Knecht1}.

\subsection{Discussion}

\begin{figure}[t!] 
\centerline{\epsfig{file=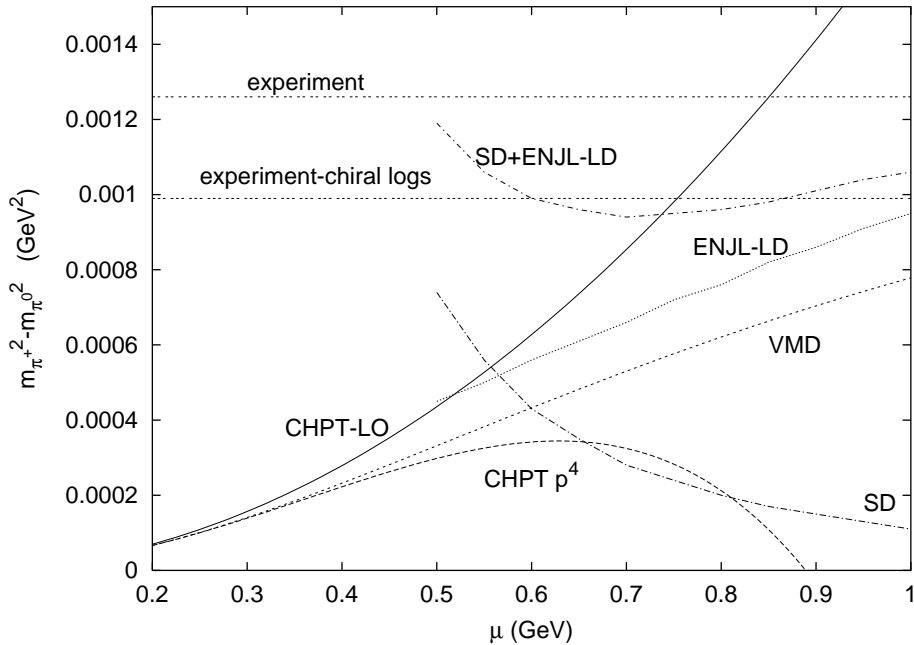,width=0.85\textwidth}}
\vspace{10pt}
\caption{The long-distance contributions and short-distance contribution
compared with the experimental values.}
\label{figdashen3}
\end{figure}

Numerical results are shown in Fig~\ref{figdashen3} for all cases.
The experimental value and the one with the sub-leading in $1/N_c$
subtracted are shown as the horizontal lines. The subtracted part
is the chiral logarithm contribution as estimated in \cite{BP2}.
Notice that CHPT starts deviating quickly from the VMD and ENJL results.
The CHPT result is only reliable up to about 500 MeV.
The VMD result and the ENJL result basically coincide here, the difference
is due to the precise input values. Both these curves also follow
essentially the correct answer as obtained from electron-positron annihilation
and the sum-rule of \cite{das}.

Notice that the ENJL model has the correct matching on to the low $\mu$
CHPT result and is a considerable improvement over it at higher $\mu$.
Notice also the almost perfect agreement with the estimated part leading
$N_c$ part of the mass-difference.

From this section we can conclude:
\begin{enumerate}
\item Different Low energy models give quite different results and
we have to use short-distance constraints and phenomenological inputs to
improve the long-distance contribution to above the regime where
CHPT is applicable. 
\item CHPT alone for the long-distance regime is as a first guestimate
acceptable but start differing from the correct answer at a scale
of about 500~MeV.
\item Even for this low-momentum dominated observable the short-distance
contributions are sizable at scales around 800~MeV.
\end{enumerate}

\section{Kaon non-leptonic Decays}
\label{kaon}

\begin{figure}[t!] 
\centerline{\epsfig{file=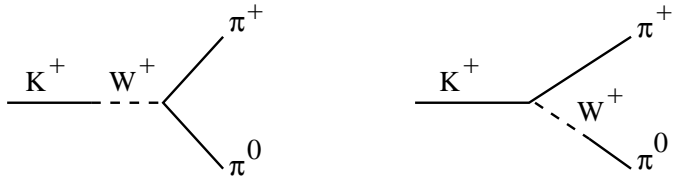,width=0.6\textwidth}}
\vspace{10pt}
\caption{The naive $W^+$ exchange contribution to $K^+\to\pi^+\pi^0$.}
\label{figkdecay}
\end{figure}

One of the difficult unresolved problems is to understand
the origin of the $\Delta I=1/2$ rule. The underlying process
is $W^+$-exchange leading to an operator of the quark-structure
$(\bar{s}u)(\bar{u}d)$ which has both isospin $1/2$ and isospin $3/2$
pieces. If we assume the $W^+$ couples directly to hadrons
the process $K^+\to\pi^+\pi^0$ goes simply via the diagrams in 
Fig.~\ref{figkdecay}, but there are no such diagrams for
$K^0\to\pi^0\pi^0$ because of charge conservation.
So we would expect that $\Gamma(K^+\to\pi^+\pi^0)\gg\Gamma(K^0\to\pi^0\pi^0)$.
The experimental numbers are
$\Gamma(K^+\to\pi^+\pi^0)=1.1~10^{-14}~\mbox{MeV}$ and
$\Gamma(K^0\to\pi^0\pi^0)
= \frac{1}{2}\Gamma(K_S\to\pi^0\pi^0)=2.3~10^{-12}~\mbox{MeV,}$
precisely the opposite.
Translated into isospin amplitudes for the decays, see e.g. \cite{BPP}
for the precise definitions, we obtain
$
\left|A_0/A_2\right|_{\mbox{exp}} = 22.1\;.
$
The problem is not due to chiral corrections since using the estimate
of \cite{BPP,KMW} we can extract them and get
\begin{equation}
\left|A_0/A_2\right|_{\mbox{chiral}} = 16.4 
\underbrace{ = \sqrt{2}}_{\mbox{naive}}\,.
\end{equation}
where the last number is the one using naive $W^+$-exchange as
depicted in Fig.~ \ref{figkdecay}.
In the notation used in \cite{BP1,BPP} we have
\begin{equation}
A_0 = C(9 G_8+G_{27})\sqrt{6}/9~F_0(m_K^2-m_\pi^2)\quad
A_2 = C 10 G_{27}\sqrt{6}/9~F_0(m_K^2-m_\pi^2)
\end{equation}
which after subtracting the estimated chiral corrections from experiment yields
\begin{equation}
G_8 = 6.2\pm0.7\quad G_{27}=0.48\pm0.06\;.
\end{equation}
Both $G_8$ and $G_{27}$ are equal to one in the $W^+$-exchange limit,
the constant $C$ was chosen to have this. We thus have to explain the large
deviation from 1 using the corrections suppressed by $1/N_c$.

This is not a hopeless task as the sub-leading corrections coming from the
diagrams in Fig.~\ref{figdashen} with the photon replaced by the gluon
are of order 
\begin{equation}
\frac{\alpha_S}{N_c} \log\frac{M_W^2}{\mu^2}
\end{equation}
compared to the leading contribution and this 
is in fact larger than one.

Luckily we know how to resum this type of logarithms\cite{buraslectures}.
At a high scale we can replace the effect of $W^+$-exchange
by a sum of local operators by virtue of the operator product
expansion. We can then use the whole renormalization group machinery
to run this sum over local four-quark operators down to a low scale $\mu_R$.
This is explained in great detail in \cite{buraslectures}.
The end result is
\begin{equation}
H_W = \sum_{i=1,10} C_i(\mu_R) Q_i
\end{equation}
with a series of known coefficients $C_i(\mu_R)$, the Wilson coefficients.
The final answer is then the
matrix element of this sum over four-quark operators,
$\langle\pi\pi|H_W|K\rangle$.

But here we have two problems:
\begin{enumerate}
\item In the previous section the short and long-distance contributions were
separated via the photon momentum. Here we have to link this somehow to
the scale $\mu_R$ appearing in the weak Hamiltonian $H_W$.
\item To next-to-leading order in the renormalization group the coefficients
$C_i(\mu_R)$ also depend rather strongly on the precise definition of the local
four quark operators $Q_i$ in QCD perturbation theory.
\end{enumerate}

In \cite{BPBK} we showed how the method of \cite{BBG} supplemented
with the correct momentum routing \cite{BBGpi,BGK} solved problem 1.
In \cite{BP1} and in the various already published talks \cite{talks} we
showed how a careful identification across the long-short-distance boundary
is also possible in this case. The basic idea is to go at the scale $\mu_R$
back from the local four-quark operators to the exchange of a series of
$X$-bosons. These $X$-bosons can then be treated in exactly the same way
as we did the photon in the previous section, thus allowing a correct
calculation at all length scales.

\begin{figure}[t!]
\centerline{\epsfig{file=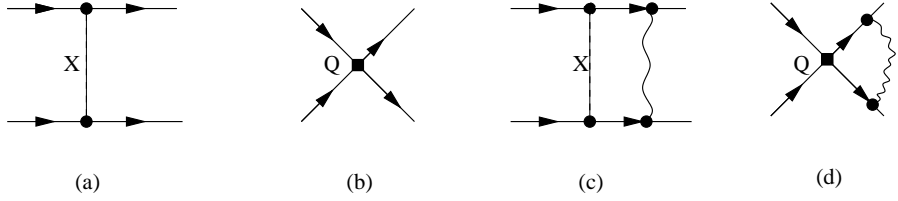,width=0.8\textwidth}}
\vskip10pt
\caption{\label{figX} The diagrams needed for the identification
of the local operator $Q$ with $X$-boson exchange in the case of
only one operator and no Penguin diagrams. The wiggly line
denotes gluons, the square the operator Q and the dashed line
the $X$-exchange. External lines are quarks.}
\end{figure}

So we replace, using a single operator as an example
\begin{equation}
C_1 Q_1 = (\bar{s}_L\gamma_\mu d_L)(\bar{u}_L\gamma_\mu u_L)
\Longleftrightarrow
X_\mu\left[g_1 (\bar s_L\gamma^\mu d_L)+g_2 (\bar u_L\gamma^\mu u_L)
\right]\,.
\end{equation}
Using the tree level diagrams of Fig.~\ref{figX}
this gives $C_1 = {g_1 g_2}/{M_X^2}$.
If we now include the one loop diagrams we obtain instead:
\begin{equation}
C_1\left(1+\alpha_S(\mu_R)r_1\right)
= \frac{g_1 g_2}{M_X^2}\left(1+\alpha_S(\mu)a_1+
\alpha_S(\mu_R)b_1\log\frac{M_X^2}{\mu_R^2}\right)\,.
\end{equation}
On the l.h.s. the scheme dependence disappears but there is a
dependence in $r_1$ on the choice of external states. 
The exact same dependence in $a_1$ cancels this. 

We now split
the integral over the $X$-boson momentum as in the previous section
\begin{equation}
\label{split}
\int_0^\infty dp_X 
\Longrightarrow
\int_0^{\mu}dp_X 
+\int_{\mu}^\infty dp_X 
\end{equation}
In the final answer all $M_X$ dependence drops out,
the logarithm
proportional to $b_1$ shows up in precisely the same way in the evaluation
of the short distance part of (\ref{split}) which is
proportional to $g_1g_2/M_X^2\left\{\alpha_S(\mu) a_2
+\alpha_S(\mu)b_1\log(M_X^2/\mu^2)\right\}$
The coefficients $r_1$, $a_1$ and $a_2$ give the corrections to the naive 
$1/N_c$-method.


We now use the $X$-boson method described above and put $\mu=\mu_R$.
The low energy part can be calculated using CHPT, this is the approach
used originally by \cite{BBG} and presently pursued by \cite{Hambye}
without including the corrections due to the change in scheme
when going to the long-distance part. Their results coincide with ourss
when we restrict our results to their approximations.
We obtain\cite{BP1}
\begin{eqnarray}
B_K = 0.6\mbox{---}0.8 &\quad& B_K^\chi = 0.25\mbox{---}0.40
\quad
G_8 = 4.3\mbox{---}0.75\nonumber\\
 G_{27} = 0.25\mbox{---}0.40&\quad&
G_8^\prime = 0.8\mbox{---}1.1 \quad B_6(\mu=0.8~\mbox{GeV})\approx2.2
\end{eqnarray}
$B_K$ is the bag-parameter relevant for $K^0-\overline{K^0}$ mixing
at the physical quark masses and  $B_K^\chi$ the same in the chiral limit.
The quark mass corrections are quite sizable.
The results for $G_8$ and $G_{27}$ are obtained without any free input and
agree within the uncertainties of the method with the experimental values.
We conclude that we basically now have a first principle understanding
of the $\Delta I=1/2$ rule. We discuss the various contributions below.
$G_8^\prime$ is the coefficient of the weak mass term, it contributes
at leading order to processes like $K_L\to\gamma\gamma$\cite{BPP} and
is often forgotten in those analyses.
Finally $B_6$ is much larger than used in all the analysises of the recent
experimental results for $\epsilon'/\epsilon$\cite{ktevna48} which is very
encouraging.

\begin{figure}[b!]
\centerline{\epsfig{file=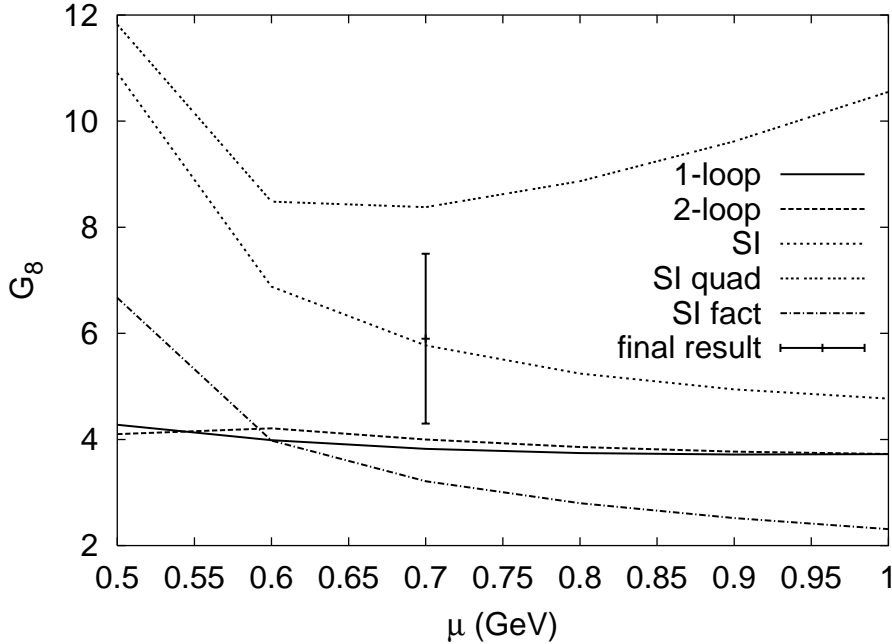,width=0.85\textwidth}}
\vspace{10pt}
\caption{The octet coefficient $G_8$ as a function of
$\mu$ using the ENJL model and the one-loop Wilson coefficients,
the 2-loop ones and those including the $r_1$ (SI). In
the latter case also the factorization (SI fact)
and the approach of 
[19] (SI~quad) are shown.}
\label{figg8}
\end{figure}

The final result for $G_8$ is depicted in Fig. \ref{figg8}.
We have shown the one-loop result (1-loop), the two-loop result with
NDR Wilson coefficients (2-loop) and the two-loop results with
correction for the long-distance scheme added (SI) using our results for the
long distance part. We also showed what naive factorization with the SI
Wilson coefficients would give and what the method of \cite{Hambye} would
give in the chiral limit with the same Wilson coefficients (SI~quad).

\begin{figure}[t!]
\centerline{\epsfig{file=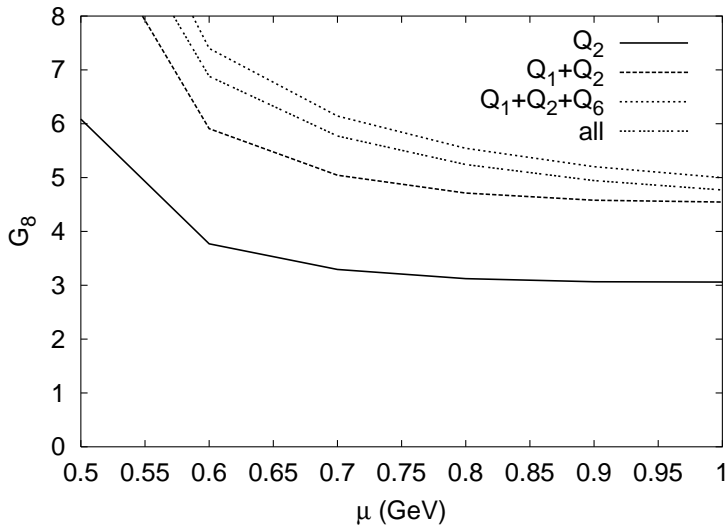,width=0.69\textwidth}}
\vspace{10pt}
\caption{\label{figg8_comp} The composition of $G_8$ as a function of
$\mu$. Shown are $Q_2$, $Q_1+Q_2$, $Q_1+Q_2+Q_6$ and all 6 $Q_i$.
The coefficients $r_1$ are included in the Wilson coefficients.}
\end{figure}

If we look at the various contributions to $G_8$ we see in
Fig.~\ref{figg8_comp} that
the contribution of $Q_1$ and $Q_2$ are both large and fairly
constant while $Q_6$ contributes 20\% or less. If we look inside the 
calculation we see that the difference with the $G_{27}$ evolution
is mainly given by the long-distance Penguin-like contributions to $Q_2$.
\begin{table}[t!]
\caption{$B_6$ as a function of $\mu$ 
using the results of 
[2].}
\label{tab:B6}
\begin{tabular}{ccddddd} 
$\mu$ (GeV) & &0.6 & 0.7 & 0.8 & 0.9 & 1.0 \\
\tableline
CHPT & &1.19 & 0.93 & 0.70 & 0.50 & 0.36 \\
ENJL & &2.27 &2.16 & 2.11&  2.11& 2.14\\
\end{tabular}
\end{table}
The behaviour of $B_6$ is more difficult, it is ill-defined in
the chiral limit in the factorizable approximation\cite{BP1} and we can thus
only define it with respect to the full large $N_c$ limit.
Calculating it in CHPT only then gives fairly low values as is visible
in the second line of Table \ref{tab:B6}. Adding higher order corrections
we immediately obtain a strongly enhanced value as is obvious from the
second line in Table \ref{tab:B6}.

\section{Conclusions}
\paragraph{} The $X$-boson method in combination with large $N_c$ arguments
allows to correctly identify quantities across theory boundaries assuming we
can identify currents across the boundary.
\paragraph{} The mass difference $m_{pi^+}^2-m_{\pi^0}^2$ is well described
by these methods with a surprisingly large short-distance contribution.
\paragraph{} The $\Delta I=1/2$ rule is now quantitatively
understood to about 30\%
with NO free input. This calculation passes {\em all} requirements usually
asked of in this context but there are many technical subtleties.
\paragraph{} $B_6\approx 2.2$ is good news for those trying to
explain the observed values of $\epsilon'/\epsilon$ within the standard model.
\paragraph{} This program has been quite successful but we need
new ideas to calculate more complex processes.

\end{document}